\documentclass{aa}  
\usepackage{graphicx}
\usepackage{amsmath}
\usepackage{txfonts}
\usepackage{url}
\usepackage{natbib}
\bibpunct{(}{)}{;}{a}{}{,} % to follow the A&A style

\begin{document}
   \title{A blind test of photometric redshifts on ground-based data}

   \author{H. Hildebrandt
          \inst{1,2}
          \and
          C. Wolf\inst{3}
          \and
          N. Ben\'itez\inst{4}
          }

   \offprints{H. Hildebrandt}

   \institute{Argelander-Institut f\"ur Astronomie, Auf dem H\"ugel 71, D-53115, Germany; \email{hendrik@astro.uni-bonn.de}
         \and
         Sterrewacht Leiden, Niels Bohrweg 2, NL-2333 CA Leiden, The Netherlands; \email{hendrik@strw.leidenuniv.nl}
         \and
             Department of Physics, University of Oxford, DWB, Keble Road, Oxford, OX1 3RH, U.K.; \email{cwolf@astro.ox.ac.uk}
             \and
             Instituto de Matem\'aticas y F\'isica Fundamental (CSIC), C/Serrano 113-bis, 28006, Madrid, Spain; \email{benitez@iaa.es}
             }

   \date{Received;  accepted}

   \abstract{}{ Several photometric redshift (photo-$z$) codes are
     discussed in the literature and some are publicly available to be
     used by the community. We analyse the relative performance of
     different codes in blind applications to ground-based data. In
     particular, we study how the choice of the code-template
     combination, the depth of the data, and the filter set influences
     the photo-$z$ accuracy.}{We performed a blind test of different
     photo-$z$ codes on imaging datasets with different depths and
     filter coverages and compared the results to large spectroscopic
     catalogues. { We analysed the photo-$z$ error behaviour to select
       cleaner subsamples with more secure photo-$z$ estimates.} We
     consider \emph{Hyperz}, \emph{BPZ}, and the code used in the
     CADIS, COMBO-17, and HIROCS surveys.}  {{ The photo-$z$ error
       estimates of the three codes do not correlate tightly with the
       accuracy of the photo-$z$'s. While very large errors sometimes
       indicate a true catastrophic photo-$z$ failure, smaller errors
       are usually not meaningful.} For any given dataset, we find
     significant differences in redshift accuracy and outlier rates
     between the different codes { when compared to spectroscopic
       redshifts}.  However, different codes excel in different
     regimes. { %The comparison of the different photo-$z$'s against
%       each other reveals a different behaviour on a subsample with
%       secure spectroscopic redshifts than on the whole
%      catalogue. 
The agreement between different sets of photo-$z$'s is better for the
subsample with secure spectroscopic redshifts than for the whole
catalogue.  Outlier rates in the latter are typically larger by at
least a factor of two.}} {{ Running today's photo-$z$ codes on
       well-calibrated ground-based data can lead to reasonable
       accuracy. The actual performance on a given dataset is largely
       dependent on the template choice and on realistic instrumental
       response curves. The photo-$z$ error estimation of today's
       codes from the probability density function is not reliable,
       and reported errors do not correlate tightly with accuracy. It
       would be desirable to improve this aspect for future
       applications so as to get a better handle on rejecting objects
       with grossly inaccurate photo-$z$'s. The secure spectroscopic
       subsamples commonly used for assessments of photo-$z$ accuracy
       may be biased toward objects for which the photo-$z$'s are
       easier to estimate than for a complete flux-limited sample,
       resulting in very optimistic estimates.}}


   \maketitle

\section{Introduction}
\label{sec:introduction}
Photometric redshifts (hereafter, photo-$z$) have become a standard
tool for the observing astronomer in the last years {
  \citep{1986ApJ...303..154L, 1995AJ....110.2655C,
    1999ASPC..191....3K, 1999A&A...343..399W, 2000ApJ...536..571B,
    2000A&A...363..476B, 2001AJ....122.1151R, 2001AJ....122.1163B,
    2001A&A...365..660W, 2003AJ....125..580C, 2003MNRAS.339.1195F,
    2004PASP..116..345C, 2004MNRAS.353..654B, 2006A&A...457..841I,
    2006MNRAS.372..565F}}.  Not only are large multi-colour imaging
surveys planned and executed with the goal of estimating the redshift
of as many galaxies as possible from their broad-band photometry, but
also many smaller projects benefit from this technique by providing
redshifts that are much cheaper, in terms of telescope time, than
spectroscopic ones and may go deeper.

{ Users of photo-$z$'s are often concerned with three main
  performance issues, which are the mean redshift error, the rate of
  catastrophic failures, and the validity of the probability density
  function (PDF) in a frequentist interpretation. The PDF may be
  correct in a Bayesian interpretation when including systematic
  uncertainties in the model fitting and correctly express a degree of
  uncertainty.  However, given the non-statistical nature of
  systematic uncertainties a frequentist PDF that correctly describes
  the redshift distribution in the real experiment is necessarily
  different, unless such systematics can be excluded.

The process depends on three ingredients: model, classifier, and
data. A basic issue at the heart of problems with the PDF are the
match between data and model, since best-fitting parameters and
confidence intervals in $\chi^2$-fitting are only reliable when the
model is appropriate.  The importance in choosing the type of data is
the need to break degeneracies between ambiguous model
interpretations. Finally, the classifiers are expected to produce
similar results, while they could produce them at dramatically
different speed. Artificial Neural Nets, hereafter ANNs, are
especially fast once training has been accomplished
\citep{2003MNRAS.339.1195F}.

There are many cases in the literature where the precision of
photo-$z$'s has been improved after recalibrating the match between
data and model \citep[see
  e.g. ][]{2003AJ....125..580C,2004ApJS..150....1B,2006A&A...457..841I,2006AJ....132..926C,2007ApJS..172...99C},
although this process requires a large, representative training set of
spectroscopic redshifts from the pool of data that is to be
photo-$z$'ed.  If ANNs are trained with sufficiently large training
samples they can achieve the highest accuracies within the training
range as a mismatch between data and model is ruled out from the
start.

The literature reports several different photo-$z$ estimators in use
across the community, some of which use different template models and
some of which allow implementation of user-defined template sets.
Assuming a modular problem, where model (templates), classifier, and
data can be interchanged, it is interesting to test how comparable the
results of different combinations are. In this spirit, we have started
the work presented in this paper, where we analyse photo-$z$
performance from real ground-based survey data, in dependence of
magnitude, depth of data, filter coverage, redshift region, and choice
of photo-$z$ code. We concentrate on the blind performance of
photo-$z$'s which is the most important benchmark for any study that
cannot rely on recalibration, e.g. in the absence of spectroscopic
redshifts. We choose to focus on ground-based datasets because a lot
of codes were tested on the Hubble-Deep-Field for which results can
already be found in the literature \citep[see
  e.g. ][]{1998AJ....115.1418H, 2000A&A...363..476B,
  2000ApJ...536..571B}.

Meanwhile, a much larger initiative has formed to investigate all
(even subtle) differences in workings and outcomes among codes and
models. This initiative called
PHAT\footnote{\url{http://www.strw.leidenuniv.nl/~hendrik/PHAT}}
(PHoto-$z$ Accuracy Testing) engages a world-wide community of
photo-$z$ developers and users and will hopefully develop our
understanding of photo-$z$'s to a reliably predictive level.}

The paper is organised as follows. In Sect.~\ref{sec:data} the imaging
and spectroscopic datasets are presented. The photo-$z$
codes used for this study are described in Sect.~\ref{sec:codes}.
Sect.~\ref{sec:strategy} presents our approach for describing
photo-$z$ accuracy. The results are presented and discussed
in Sect.~\ref{sec:results}. { The different photo-$z$ estimates are
  compared to each other in Sect.~\ref{sec:phz_vs_phz}.} A final
summary and general conclusions are given in
Sect.~\ref{sec:conclusions}.

Throughout this paper we use Vega magnitudes if not otherwise
mentioned.

\section{Datasets}
\label{sec:data}
We investigate the performance of photo-$z$'s on three
different imaging datasets:

\begin{enumerate}
\item 
We use five-colour $UBVRI$ data from the ESO Deep Public Survey (DPS)
field Deep2c centred on the Chandra Deep Field South (CDFS) which were
observed with the Wide Field Imager (WFI) at the 2.2m telescope at La
Silla, Chile, reduced with the THELI reduction pipeline, and described
in detail in \cite{2006A&A...452.1121H} as part of the Garching-Bonn
Deep Survey (GaBoDS). The data originate from different projects,
mostly from the ESO Imaging Survey (EIS), the COMBO-17 survey, and the
GOODS program.  Results for these data can be regarded as
representative for very deep ground-based, wide-field surveys with the
typical photometric accuracy achievable for multi-chip camera,
multi-epoch data.

In order to measure unbiased object colours the $BVRI$ images were
filtered to the seeing of the $U$-band ($\approx1\farcs0$) and the
photometric catalogue was created with \emph{SExtractor}
\citep{1996A&AS..117..393B} in dual-image-mode with the unfiltered
$R$-band image as the detection image.

\item
On the same field and taken with the same camera there are catalogues
available from the COMBO-17 survey covering the same broad-band
filters in $BVRI$ to considerably shallower depth, a different
$U$-band filter, and 12 additional medium-band filters in the optical
wavelength range.  These data are described in detail in
\cite{2004A&A...421..913W}. In terms of exposure time the COMBO images
are shallower by a factor of 2.5 ($R$-band) to 12.5 ($V$-band)
corresponding to approximately 0.4-1.7 magnitudes.

The broad-band data from COMBO-17 resemble a medium-deep wide-field
survey, while the full 17-filter data are presently unique in its
kind.  However, we can use them to investigate whether additional
telescope time should be spent on increasing depth as in GaBoDS or on
obtaining additional SED information as in COMBO-17.

In contrast to GaBoDS, the COMBO-17 photometry was measured directly
on unfiltered images. The photometry was obtained in Gaussian
apertures whose width was adapted to compensate seeing variations
between the frames. Provided the convolution of aperture and PSF
yields the same result for each frame, this procedure is
mathematically identical to filtering all frames to a final constant
seeing and extracting fluxes with Gaussian apertures at the end.

{ The calibration of the CDFS field of COMBO-17 has however changed
  since the original publication of the data in 2004. COMBO-17 is
  calibrated by two spectrophotometric standard stars in each of its
  fields. However, the two stars on the CDFS suggested calibrations
  that were inconsistent in colour at the 0.15~mag level from $B$ to
  $I$.  Both were marginally consistent with the colours of the
  Pickles atlas, so the choice was
  unconstrained. \citet{2004A&A...421..913W} ended up trusting the
  wrong star and introducing a colour bias towards the blue. The
  calibration has since been changed to follow the other star, and is
  now consistent with both the GaBoDS and MUSYC (Multiwavelength
  Survey by
  Yale-Chile\footnote{\url{http://www.astro.yale.edu/MUSYC/}})
  calibration. The consequences of the calibration change for the
  photo-$z$'s is little in the 17-filter case, but large when only
  using broad bands. Broad-band photo-$z$'s hinge more on colours than
  on features that are traced in medium-band photo-$z$'s.}

\item
Furthermore, we use catalogues from the FORS Deep Field \citep[FDF;
  see][]{2003A&A...398...49H, 2004Gabasch_PhD_Thesis} involving eight
broad-band filters, $UBgRIZJK_s$, observed with FORS@VLT in the
optical and SOFI@NTT in the near-infrared. At least in the optical,
these data are representative of very deep pencil-beam surveys
achievable with present day large telescopes. { With this dataset
  we are able to quantify the impact of adding near-infrared data to
  deep optical data on photo-$z$'s.} The FDF photometric catalogue
contains flux measurements in apertures of different sizes obtained
after filtering images to the same PSF. In the following, we use
fluxes in aperture diameters of $d=1\farcs5$ .

\end{enumerate}

\subsection{Comparisons of imaging data}
\label{sec:comparisons}
In all three datasets, the multi-band fluxes of a given object were
effectively measured in identical physical apertures outside the
atmosphere (and with identical spatial weighting) for all filters,
assuming that seeing produces a Gaussian-shaped PSF.  Colours could
still be biased by non-Gaussianity of the PSF and by suboptimal
background subtraction.

The properties of these three imaging datasets are summarised in
Table~\ref{tab:imaging}. Since the limiting magnitudes are estimated
in completely different ways in the three data release papers, we
decided to calculate hypothetical 10$\sigma$ limiting magnitudes 
with the GaBoDS values as a reference. These $m_{\mathrm{lim,eff}}$ 
correspond to the 10$\sigma$ sky noise under the following assumption:
\begin{equation}
m_{\mathrm{lim,eff,X}}-m_{\mathrm{lim,G}}=-2.5 \log \left(\frac{\mathrm{FWHM_{\mathrm{X}}}}{\mathrm{FWHM_{\mathrm{G}}}}\right)\sqrt{\frac{t_{\mathrm{exp,G}}}{t_{\mathrm{exp,X}}}}\left(\frac{2.2~{\rm m}}{D}\right)\,,
\end{equation}
with $\mathrm{G}$ denoting GaBoDS quantities and $\mathrm{X}$ denoting
quantities of the other dataset. FWHM is the measured seeing,
$t_{\mathrm{exp}}$ is the exposure time, and $D$ is the diameter of
the telescope. By doing so we neglect differences between similar
filter transmission curves and variations in observing conditions
(moon, sky transparency etc.) Thus, the limiting magnitudes
are only rough estimates for approximate comparison.

The FDF limiting magnitudes in the $ZJKs$-bands are the ones given in
\cite{2003A&A...398...49H} and \cite{2004Gabasch_PhD_Thesis}
corresponding to 50\% completeness.

\begin{figure*}
%\sidecaption
\includegraphics[width=12cm]{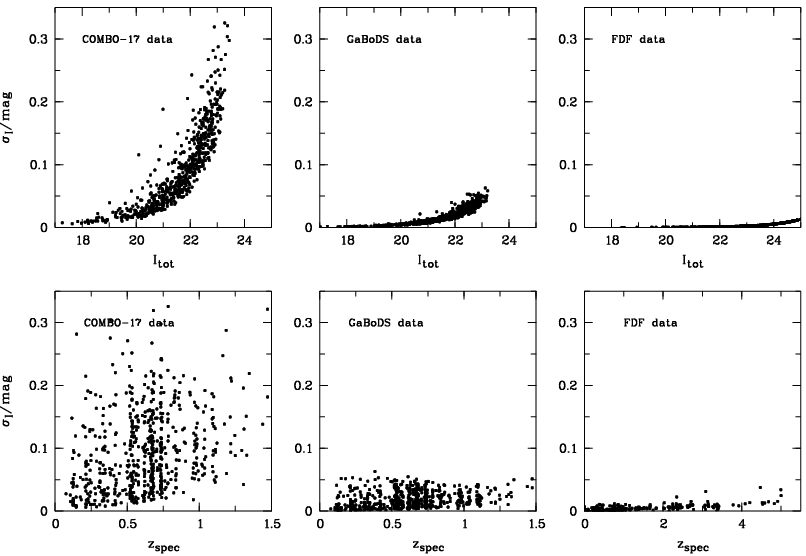}
\caption{\label{fig:errors}Photometric errors in the $I$-band as a
  function of $I$-magnitude (\emph{upper panel}) and as a function of
  spectroscopic redshift (\emph{lower panel}) for the COMBO-17 data
  (\emph{left}), the GaBoDS data (\emph{middle}), and the FDF data
  (\emph{right}); see text for information on how the errors were
  estimated. { Note that the errors of the photometric zeropoints
    are larger than the purely statistical errors plotted here.}  }
\end{figure*}

The dependence of photometric errors on magnitude and redshift in the
three datasets is shown in Fig.~\ref{fig:errors}. The errors for the
COMBO data are derived from multiple measurements of the same sources,
where photon shot-noise is assumed to be a lower limit.  The GaBoDS
and FDF errors are purely derived from shot-noise as no multiple 
measurements were made.

\begin{figure*}
%\sidecaption
\includegraphics[width=12cm]{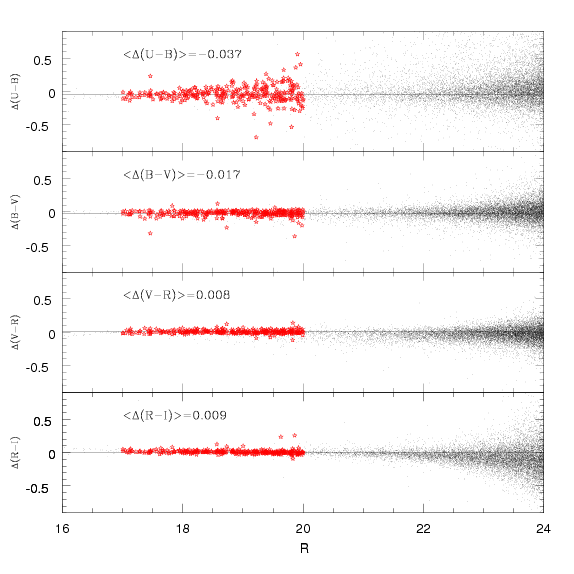}
\caption{\label{fig:colour_comp}{ Comparison of colour measurements
    for objects in the GaBoDS and the COMBO catalogues. Note that the
    $U$-band filters in the two datasets are different, with the
    GaBoDS filter being broader and bluer. The star symbols represent
    objects selected by the \emph{SExtractor} CLASS\_STAR parameter in
    the magnitude range $17<R<20$.}}
\end{figure*}

{ We compare the colour measurements in the COMBO and the GaBoDS
  catalogue and find very good agreement (see
  Fig.~\ref{fig:colour_comp}). Thus, the different ways of correcting
  for the PSF variations from band to band deliver consistent
  results. We carried out another comparison between the COMBO data
  and the CDFS catalogue from the MUSYC collaboration (E.~Taylor,
  private communication) and the agreement is similar. We conclude
  that the colour measurement cannot be a dominant source of
  systematic error in the following.}

\begin{table*}
\begin{minipage}[t]{\textwidth}
  \caption{Properties of the imaging data.}
\label{tab:imaging}
\renewcommand{\footnoterule}{}  % to avoid a line before footnotes
\begin{tabular}{c|rrc|rrc|rrc}
  \hline
  \hline
  & \multicolumn{3}{|c|}{COMBO-17} & \multicolumn{3}{|c|}{GaBoDS} & \multicolumn{3}{|c}{FDF}\\
  Band & Exp. time\footnote{The total exposure time of the COMBO-17 medium-bands on the CDFS is 108~ksec.} & FWHM & $m_{\mathrm{lim,eff}}$ [mag]\footnote{Effective limiting magnitudes calculated as described in the text} & Exp. time & FWHM &  $m_{\mathrm{lim}}$ [mag] & Exp. time & FWHM &  $m_{\mathrm{lim,eff}}$ [mag]$^b$ \\
       &       [s] &      &  10$\sigma$ sky noise & [s] & & 10$\sigma$ sky noise & [s] & & 10$\sigma$ sky noise\\
  \hline
  $U$         & $21\,600$ & $1\farcs00$ & 24.6 &  $78\,900$ & $1\farcs01$ & 25.3 & $44\,400$ & $0\farcs97$ & 26.4\\
  $B$         & $11\,240$ & $1\farcs10$ & 26.2 &  $69\,400$ & $0\farcs98$ & 27.3 & $22\,700$ & $0\farcs60$ & 28.6\\
  $V/g$       &    $8400$ & $1\farcs20$ & 25.2 & $104\,600$ & $0\farcs92$ & 26.9 & $22\,100$ & $0\farcs87$ & 27.5\\
  $R$         & $35\,700$ & $0\farcs75$ & 26.4 &  $87\,700$ & $0\farcs79$ & 26.8 & $26\,400$ & $0\farcs75$ & 27.6\\
  $I$         &    $9800$ & $1\farcs20$ & 23.6 &  $34\,600$ & $0\farcs93$ & 24.6 & $24\,900$ & $0\farcs53$ & 26.4\\
  $Z$         &         - &           - &    - &          - &           - &    - & $18\,000$ & $0\farcs48$ & 25.3\\
  $J$         &         - &           - &    - &          - &           - &    - &      4800 & $1\farcs20$ & 22.9\\
  $K_s$       &         - &           - &    - &          - &           - &    - &      4800 & $1\farcs24$ & 20.7\\
  \hline
\end{tabular}
\end{minipage}
\end{table*}

\subsection{Spectroscopic catalogues}
\begin{figure}
\resizebox{\hsize}{!}{\includegraphics[angle=-90]{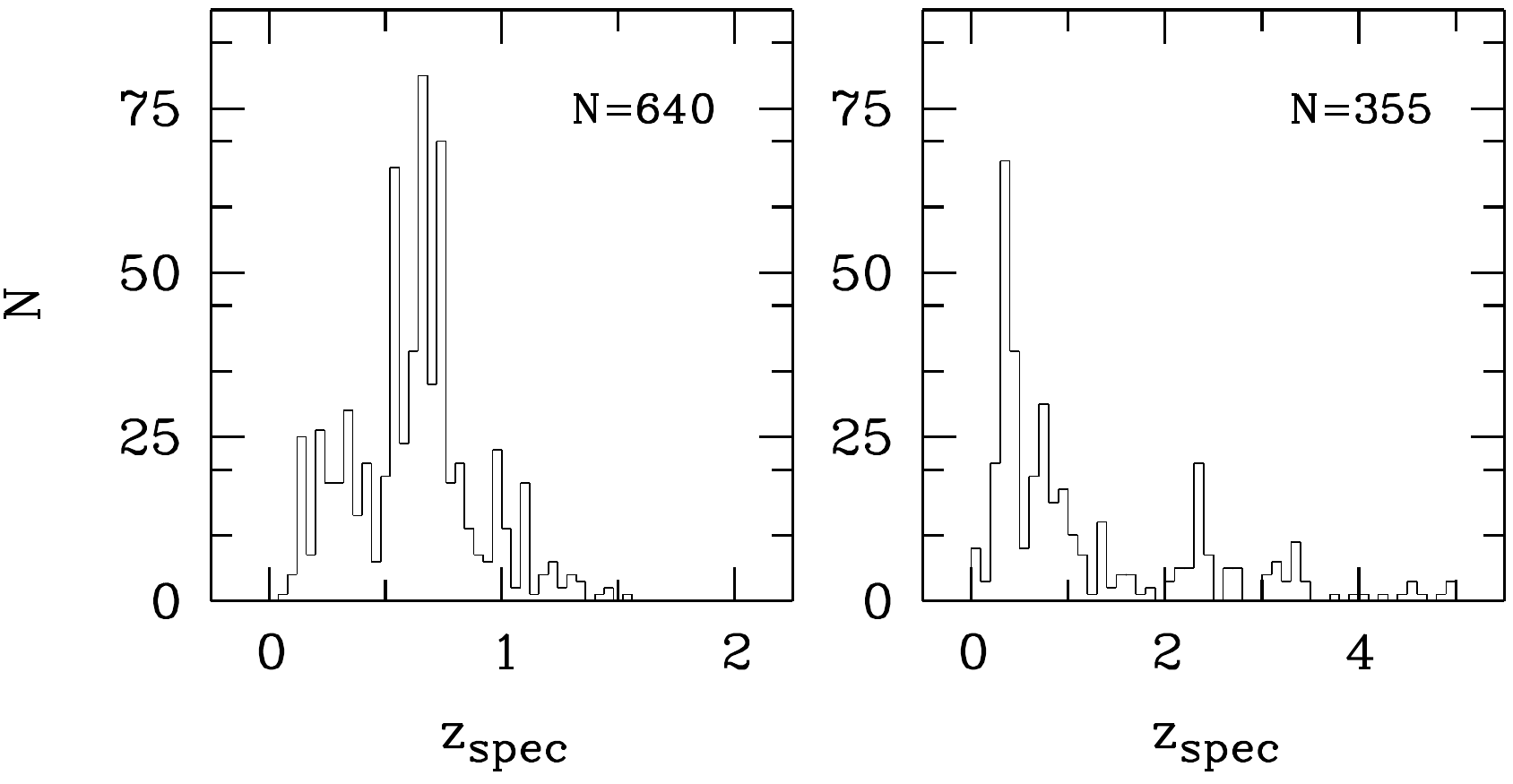}}
\caption{\label{fig:z_dist}Spectroscopic redshift distributions of the
  comparison samples. {\it Left:} The VVDS-CDFS spectroscopic data with
  $R_{\mathrm{WFI}}<24$ and flags 3,4,23,24. 
  {\it Right:} The FDF spectroscopic data.}
\end{figure}

Spectroscopic catalogues are publicly available for both fields:

\begin{enumerate}
\item The VIMOS VLT Deep Survey (VVDS) team carried out an $I_{AB}<24$
  magnitude limited spectroscopic survey on the CDFS with VIMOS@VLT
  \citep{2004A&A...428.1043L} yielding 1599 redshifts including 1452
  galaxies. The redshift measurements have associated reliability
  flags, and in the following comparisons we use only objects with
  flags 3 or 4 (or secondary targets with flags 23 or 24) indicating
  95\% and 100\% confidence, respectively, to avoid errors introduced
  by the spectroscopic catalogue. This leaves us with 640 objects with
  $R_{\mathrm{WFI}}<24$, whose redshift distribution is shown in
  Fig.~\ref{fig:z_dist}. Since there are very few objects beyond
  redshift $z\approx1.2$ this catalogue is ideally suited to assess
  the performance of photo-$z$'s with optical data alone.

\item The FDF team measured the redshifts of 355 objects with FORS@VLT
  \citep[341 of which are published in][]{2004A&A...418..885N}
  pre-selected by photo-$z$'s to cover the range $0<z<5$. The
  spectroscopic redshift distribution is also shown in
  Fig.~\ref{fig:z_dist} in comparison to the one of the VVDS
  data. This deep dataset extends well beyond the region where optical
  photo-$z$'s work well and can illustrate the benefit of
  near-infrared data on photo-$z$ performance for $z>1$.

\end{enumerate}

\section{Photo-$z$ codes}
\label{sec:codes}
There are basically two different approaches to estimate a photo-$z$
for a galaxy, the ``SED-fitting''-method and the ``empirical training
set''-method. The former relies on a sample of synthetic or observed
spectral-energy-distributions (SEDs) and on theoretical knowledge how
those SEDs evolve with redshift. The latter relies on a colour
catalogue of spectroscopically observed galaxies as large as possible
to cover essentially every galaxy type at all redshifts. See e.g.
\cite{2000ApJ...536..571B} for a detailed review of both techniques
and their differences.

The empirical approach can lead to very precise results if an
extensive, complete spectroscopic catalogue with colour information is
available. But it is not as flexible as the ``SED-fitting'' method
because for every new filter set or camera the colour catalogue must
be recreated. { Moreover, it is essentially limited to the
  magnitude range where spectra are available in large numbers
  ($I\la24$) and implicit priors are driven by the spectroscopic
  sample selection.}

The ``SED-fitting''-method, however, can be applied to every dataset
for which the filter transmission curves are known. { Since we want
  to give guidance for blind applications of photo-$z$'s here we
  concentrate on this approach in the following.}

In practise, a photo-$z$ analysis often involves aspects from both
approaches. Empirical colour-redshift relations can certainly be
extrapolated in magnitude or redshift. Also, a spectroscopic catalogue
can help to optimise parts of an ``SED-fitting'' approach.
\cite{2006A&A...457..841I}, e.g., present a method to improve the
photo-$z$ estimates in the CFHT Legacy Survey. They adjust
the photometric zeropoints of their images and optimise the template
SEDs with help of more than 3000 spectroscopically observed galaxies
in the range $0<z<5$. The optimisation of templates was already used
for improving template based photo-$z$ estimates in the SDSS
\citep{2003AJ....125..580C}. \cite{2004A&A...421...41G} claim to
obtain highly accurate photo-$z$'s in the FDF by
constructing semi-empirical template SEDs from 280 spectroscopically
observed galaxies in the FDF and the Hubble Deep Field.  In the
following we describe the three codes used for this study.

\subsection{\emph{Hyperz}}
\label{sec:hyperz}
The ``SED-fitting'' photo-$z$ code \emph{Hyperz}
\citep{2000A&A...363..476B} is publicly
available\footnote{\url{http://webast.ast.obs-mip.fr/hyperz/}}, well
documented and widely used by the community. For detailed information
on the code see the manual at the website or the reference paper
mentioned above.

\emph{Hyperz} comes with two different template SED sets, the mean
observed spectra of local galaxies by \cite{1980ApJS...43..393C},
hereafter CWW, and synthetic spectra created from the spectral
evolution library of \cite{1993ApJ...405..538B}, hereafter BC. We use
the BC templates for \emph{Hyperz} since for all tested setups
performance with the CWW templates is worse.  Different reddening laws
are implemented to account for the effect of interstellar dust on the
spectral shape. By default we use the reddening law of
\cite{2000ApJ...533..682C} derived for local star-forming galaxies.
The damping of the Lyman-$\alpha$-forest increasing with redshift is
modelled according to \cite{1995ApJ...441...18M}.  Another important
option influencing performance strongly is the application of a prior
on the absolute magnitude. For a given cosmology the absolute
magnitude of an object is calculated from the apparent magnitude in a
reference filter for every redshift step. The user can specify limits
to exclude unrealistically bright or faint objects. In the following
we assume a $\Lambda$CDM cosmology ($\Omega_\Lambda=0.7$,
$\Omega_\mathrm{m}=0.3$, $H_0=70\mathrm{\frac{km}{s\cdot Mpc}}$) and
allow galaxies to have an absolute $I$-band magnitude of $M_*-2.5 <
I_\mathrm{abs} < M_*+2.5$ using the local SDSS-value of
$M_{*,\mathrm{AB}}=-21.26$ from \cite{2001AJ....121.2358B}.

Besides reporting the most probable redshift estimate as a primary
solution \emph{Hyperz} can also store the redshift probability
distribution giving the probability associated with the $\chi^2$-value
for every redshift step. Furthermore, the width of this distribution
around the primary solution is provides a confidence interval, which
allows the user to identify objects with very uncertain estimates.

{ We choose a minimum photometric error of 0.1mag for \emph{Hyperz}
  to avoid unrealistically small errors in some of the bands.}

\subsection{COMBO-17 code}
\label{sec:combo_code}
The photo-$z$ code of COMBO-17 { (also used for CADIS and HIROCS;
  ``COMBO code'' hereafter)} performs two simultaneous tasks: it
classifies objects into stars, galaxies, QSOs, and white dwarfs based
on their colours, and for galaxies and QSOs it also estimates
redshifts. Here, we used a setup forcing the galaxy interpretation in
order to better compare the results to the other codes which assume a
priori that all objects are galaxies. The code is currently not
publicly available.

{ It uses a 2D age $\times$ extinction grid of templates produced
  with the PEGASE population synthesis code
  \citep{1997A&A...326..950F} and an external SMC reddening law.} For
all template details we refer the reader to
\cite{2004A&A...421..913W}. No explicit redshift-dependent prior is
used, however, for the shallow purely optical datasets of COMBO-17 and
GaBoDS, only galaxy redshifts up to 1.4 are considered, while for the
FDF dataset the whole range from $z=0$ to $z=7$ is allowed.

{ The SED fitting is done in colour space rather than in magnitude
  space. Similar to \emph{Hyperz} a lower error threshold is applied
  (0.05mag) but here for the colour indices.}

The code determines the redshift probability distribution $p(z)$ and
reports the mean of this distribution as a Minimum-Error-Variance (MEV)
redshift and its RMS as an error estimate. The code also tests the
shape of $p(z)$ for bimodality, and determines redshift and error from
the mode with the higher integral probability \citep[for all details 
see][]{2001A&A...365..660W}.

\subsection{BPZ}
\label{sec:bpz}
BPZ (Bayesian photo-$z$'s) is a public
code\footnote{\url{http://acs.pha.jhu.edu/~txitxo/}}, which implements
the method described in \cite{2000ApJ...536..571B}. It is an SED
fitting method combined with a redshift/type prior, $p(z,T|m)$, which
depends on the observed magnitude of the galaxies.  It originally used
a set of 6 templates formed by the 4 CWW set and two starburst
templates from \cite{1996ApJ...467...38K} which were shown to
significantly improve the photo-$z$ estimation. It should be stressed
that the extrapolation to the UV and IR of the optical CWW templates
used by \emph{BPZ} is quite different from the one used by
\emph{Hyperz}. The template library has been calibrated using a set of
HST and other ground based observations as described in
\cite{2004ApJS..150....1B}. This template set has been shown to
remarkably well represent the colours of galaxies in HST observations,
to the point of being able to photometrically calibrate the NIC3
Hubble UDF observations with a 0.03 magnitude error as shown in
\cite{2006AJ....132..926C}. In the latter paper two additional, very
blue templates from the Bruzual \& Charlot library were introduced, so
the current \emph{BPZ} library contains 8 templates.

The redshift likelihood is calculated by \emph{BPZ} in a similar way
as by \emph{Hyperz} minimising the $\chi^2$ of observed and predicted
colours. However, in contrast to \emph{Hyperz} no reddening is applied
to the templates relying on the completeness of the given set. After
the calculation of the likelihood, Bayes theorem is applied
incorporating the prior probability. The actual shape of this prior is
dependent on template type and $I$-band magnitude and was derived from
the observed redshift distributions of different galaxy types in the
Hubble Deep Field. By applying this prior the rate of outliers with
catastrophically wrong photo-$z$ assignments can be reduced. For
details on the procedure see \cite{2000ApJ...536..571B}.

BPZ has been extensively used in the ACS GTO program, the GOODS and
COSMOS surveys and others.

\section{Description of photo-$z$ quality}
\label{sec:strategy}
The performance of one particular setup is characterised by some basic
quantities which are described in the following. 

The mean, $\delta_z$, and the standard deviation, $\sigma_z$, of the
following quantity are calculated:

\begin{equation}
  \Delta z=(z_{\mathrm{phot}}-z_{\mathrm{spec}})/(1+z_{\mathrm{spec}})\,.
\end{equation}
Iteratively 3$\sigma$ outliers are rejected and after convergence
their fraction is given by $f_{3\sigma}$. By doing so, the outlier
fraction $f_{3\sigma}$ is not independent of the scatter $\sigma_z$.
Therefore, we additionally report the quantity $f_{0.15}$ which is the
fraction of objects for which $\Delta z>0.15$.

\subsection{Rejection of uncertain objects}
\label{sec:rejection}
As described above every photo-$z$ code gives a confidence estimate
for each object. \emph{Hyperz} and the COMBO code report confidence
intervals on the redshift while \emph{BPZ} uses the ODDS parameter.
It is obvious that an end-user will reject objects that clearly have
uncertain photo-$z$ estimates, although it is a-priori unclear how to
define these objects. Since the codes do not estimate confidence
measures in identical ways it is not possible to apply a universal
threshold.  We can get an idea of appropriate thresholds for the
different codes by varying the cuts on the confidence intervals or the
ODDS parameter, respectively. Thus, we see how the quantities
$\delta_z$, $\sigma_z$, and $f_{3\sigma}$ change with the completeness
of the remaining sample.

When using \emph{Hyperz} and the COMBO code all objects with a
probability vs.  redshift distribution that is too wide are rejected
by the following criterion:

\begin{equation}
\label{eq:completeness}
  \sigma>A\times(1+z_{\mathrm{phot}})\,,
\end{equation}
with $\sigma$ being the half-width of the 68\% confidence interval and
$A$ the parameter that is varied from 0 to 1. The fraction of rejected
objects is then called $r_A$ and the completeness then becomes
$\mathrm{compl.}=1-r_A$.

In \emph{BPZ} we reject all objects with:
\begin{equation}
\label{eq:completeness_ODDS}
  ODDS<A\,,
\end{equation}
with $A$ varied from 100\% to 0\%. The ODDS parameter put out by
\emph{BPZ} does not allow to vary the completeness over a large
interval since a lot of objects are assigned an ODDS value of 1.

\begin{figure}
\resizebox{\hsize}{!}{\includegraphics[angle=-90]{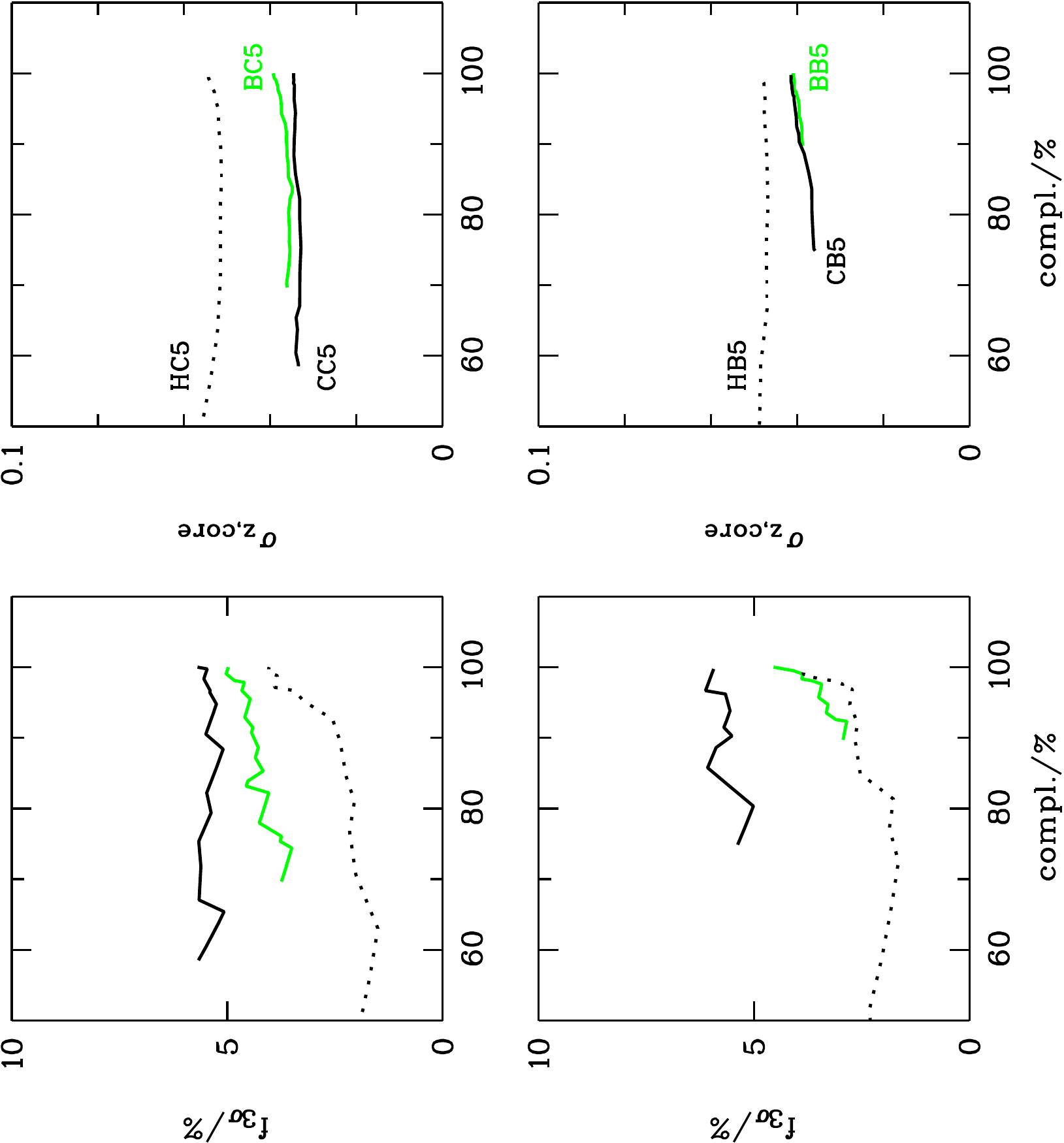}}
\caption{\label{fig:char_line}Characteristic lines showing
  completeness vs. 3$\sigma$ outlier rate, $f_{3\sigma}$,
  (\emph{left}) and vs. scatter, $\sigma_z$, (\emph{right}) for the
  COMBO (\emph{top}) and the GaBoDS (\emph{bottom}) $UBVRI$ imaging
  dataset { ($17<R<23$; \emph{BPZ}: solid grey line, COMBO code:
    solid black line, \emph{Hyperz}: dotted line).}}
\end{figure}

In this way diagrams showing $\delta_z$, $\sigma_z$, and $f_{3\sigma}$
vs. completeness are created. While $\delta_z$ is almost independent
of completeness the dependencies of $\sigma_z$ and $f_{3\sigma}$ on
completeness for selected setups are shown in Fig.~\ref{fig:char_line}.

{ Investigating many of these characteristic lines, we find that
  the most obvious feature is that $\sigma_z$ as well as $f_{3\sigma}$
  are often insensitive to a tightening of the cut criterion. This
  immediately tells us that the errors from the photo-$z$ codes are
  not proportional to the real errors on an object-by-object basis and
  thus of limited use. The real accuracy of the photo-$z$ is not
  tightly correlated with the error estimate.

The curves corresponding to \emph{Hyperz} (dotted lines) show {
  some} dependence of the outlier rate, $f_{3\sigma}$, on a tightening
of the cut. At some point around 80\% completeness a saturation
behaviour sets in and a further tightening does not decrease the
outlier rates anymore. \emph{BPZ} shows a similar but less pronounced
behaviour. Thus, very large confidence intervals or very low ODDS
values indicate that the photo-$z$ estimation failed indeed. We assume
that at this point the width of the confidence interval is not
dominated by the photometric errors but becomes influenced by
systematic uncertainties in the photometric calibration, the template
set, the filter curves or the code itself.}

From the preceding paragraphs it should be clear that the choice of
$A$ in Eq.~\ref{eq:completeness} and Eq.~\ref{eq:completeness_ODDS} as
a criterion for a reliable redshift estimate is somewhat arbitrary.
After careful investigation of all characteristic line plots for all
setups we decided to fix the cut for the rejection of uncertain
objects in \emph{Hyperz} at $\sigma>0.125$, in the COMBO code at
$\sigma>0.15$ and in \emph{BPZ} at $ODDS<0.95$. This appears to
eliminate the most uncertain objects in the datasets studied
here. { Furthermore, the error distribution of these remaining,
  secure samples is close to a Gaussian for most setups if 3-sigma
  outliers are rejected, i.e. $\sim 68\%$ of the objects lie within
  their error estimate around the mean $\delta_z$.}

There is clearly some amount of degeneracy between the quantities
defined in this section. If the photo-$z$ error distribution was
purely Gaussian, scatter and bias would be sufficient numbers to
characterise the accuracy of one particular setup. As described above,
this is not the case for real data (see also
Fig.~\ref{fig:zz_BCH_C5}~\&~\ref{fig:zz_BCH_B5}).  Usually, there is a
core which might be offset by some bias and there are very extended
wings containing catastrophic outliers. This complex error
distribution is not easily described by a few numbers and a specific
choice must be a compromise between clarity and degeneracy.

For example, a smaller core scatter will probably produce more
$3\sigma$ outliers than a larger core scatter.  With no alternative at
hand to condense the performance of one particular setup into a
handful of numbers we can only refer to the $z_{\mathrm{phot}}$ vs.
$z_{\mathrm{spec}}$ plots shown in the following which give an
uncompressed view of the data.

\subsection{VVDS setups}
The VVDS is complete {\b in terms of observations} down to
$I_\mathrm{AB}=24$ which corresponds to
$I_\mathrm{Vega}\approx23.5$. Thus, we decided to asses the photo-$z$
accuracy for all objects with $17<R<23$ to achieve a reasonable level
of completeness. { However, we note that the fraction of VVDS
  objects with high-quality flags (3/4/23/24) has dropped to
  approximately $\sim2/3$ compared to the whole VVDS catalogue at
  $R<23$.}  We also present the results for a fainter magnitude bin
with objects in the range $23<R<24$ which are then possibly biased in
terms of selection.  The mean redshifts in the bright and the faint
bins are $z=0.55$ and $z=0.75$, respectively.

The different setups are named with three-letter acronyms with the
first letter denoting the code (``H'' for \emph{Hyperz}, ``C'' for the
COMBO code, and ``B'' for \emph{BPZ}), the second letter denoting the
dataset (``B'' for GaBoDS and ``C'' for COMBO), and the digit at the
third position denoting the filter set (``5'' for $UBVRI$, ``4'' for
``BVRI'', and ``17'' for the full COMBO-17 filter set including
medium-band-filters).

\subsection{FDF setups}
Since the FDF data are extremely deep and subtle trends in photometric 
errors make little difference to the photo-$z$ quality, we do not split 
the FDF sample into magnitude bins. 
Furthermore, given the selection choices made for the spectroscopic 
sample it is not complete at or representative for any particular 
magnitude limit. Hence, we split the FDF spectroscopic catalogue 
into two samples at $z=2$ to show the effects of different filter 
sets and especially NIR bands on the performance at low and high 
redshift in comparison. The FDF setups are denoted by a second
letter ``F'' for FDF and the filter set is spelled out.

\section{Results and Discussion}
\label{sec:results}
In the following we report the results from our blind test of
photo-$z$ performance for the different setups. As a complete coverage
of all possible data-code-parameter combinations would be beyond the
scope of this paper we concentrate on some well-chosen setups to
illustrate the effects of key parameters.

\subsection{VVDS results}
The statistics for all photo-$z$ setups that are compared to the VVDS
spectroscopic catalogue are shown in Table~\ref{tab:res_VVDS_COMBO}
and Table~\ref{tab:res_VVDS_GaBoDS}.  Selected setups are also
illustrated in Fig.~\ref{fig:zz_BCH_C5} and Fig.~\ref{fig:zz_BCH_B5}
by plots showing photo-$z$ versus spectroscopic redshift.

\subsubsection{COMBO data}
Clearly, the COMBO code performs best in comparison to the two other
codes with the 17-filter set as well as with the 4- and 5-filter sets.
{ While \emph{BPZ} produces similar outlier rates and scatter
  values as the COMBO code the completeness is lower. \emph{Hyperz}
  performs slightly worse here.

\emph{BPZ} and \emph{Hyperz} produce { some} negative biases. The
cross-calibration between templates and photometry is obviously more
accurate for the PEGASE templates used by the COMBO code than for the
CWW, Kinney, and BC templates used by \emph{BPZ} and \emph{Hyperz},
respectively. Similar negative biases are found by
\cite{2003AJ....125..580C} using the CWW and BC templates for
photo-$z$ estimates on SDSS data.}

The COMBO code shows the expected behaviour that the photo-$z$
accuracy decreases when further filters are excluded. Completeness
decreases while outlier rate and scatter increase. No large biases are
produced in any setup.

{ A very interesting fact concerning \emph{Hyperz} and \emph{BPZ}
  is that the exclusion of the $U$-band \emph{decreases} the bias in
  the bright bin. Clearly, the photo-$z$ results with the COMBO code
  as well as the comparisons between the different datasets in
  Sect.~\ref{sec:comparisons} show that this behaviour is not caused
  by a badly calibrated $U$-band.}

{ The best results in both magnitude bins are certainly achieved
  with the full 17-filter set of COMBO-17. Especially in the bright
  magnitude bin the scatter and the outlier fractions are very small
  compared to all 4- or 5-filter-setups. In the fainter bin, however,
  the difference is not as dramatic due to the lack of depth in many
  of the medium-bands. \emph{Hyperz} also shows relatively accurate
  results for the 17-filter set (HC17) but not as accurate as the
  CC17. BC17 performs in between. In the bright bin, the proper
  modelling of emission lines in the PEGASE templates that can affect
  the flux in the medium-band filters considerably pays off for the
  COMBO code resulting in a very small scatter on the 0.02
  level. Emission lines are not included in the BC93 templates used by
  \emph{Hyperz} and less pronounced in the observed CWW + Kinney
  templates of \emph{BPZ}.}

\subsubsection{GaBoDS}
Owing to their greater depth the GaBoDS data mostly lead to better
results than the shallower COMBO data, in $UBVRI$ as well as in
$BVRI$. As expected, the effect is much more pronounced in the faint
bin, while the depth helps less in the estimation of redshifts for
high $S/N$ objects at the bright end of our catalogue.  Nevertheless,
also bright objects with $>20\sigma$ detections in the $R$-band can
benefit from the depth in the other bands.

The negative biases in the photo-$z$ estimation with \emph{BPZ} and
\emph{Hyperz} is also present in GaBoDS setups with the $U$-band
included. At this point, it is important to mention again that the
GaBoDS $U$-band filter is different from the COMBO $U$-band
filter. The GaBoDS filter is wider and bluer.

For the COMBO code, the 4- and 5-filter results are nearly
indistinguishable. Only in the faint bin the outlier rates increase
slightly when the $U$-band is excluded. \emph{Hyperz} shows the
unexpected feature that most statistics become more accurate when
going from five to four filters. \emph{BPZ} shows a similar behaviour
as the COMBO code. The statistics are nearly independent on the choice
between 4- and 5-filter set. Even the bias of $\sim0.06\,\mathrm{mag}$
in the faint bin is this time present when using just $BVRI$.

{ The biases for the $UBVRI$ setups may well be due to the very
  blue $U$-band filter used for the GaBoDS data. The filter-curve
  entering the photo-$z$ code is less well defined because of the
  strongly varying spectral throughput of the atmosphere in the
  near-UV and the large chip-to-chip variations in differential CCD
  efficiency at these wavelengths. We tried to shift the blue-cutoff
  of the transmission curve of the atmosphere in a reasonable
  range. This can slightly reduce the photo-$z$ bias but might not be
  reproducible. This problem is also present in the COMBO data but
  less severe due to the redder COMBO $U$-band.}

\subsubsection{Common trends}
{ The outlier rates, $f_{3\sigma}$ produced by \emph{Hyperz} are in
  most cases larger than the outlier rates produced by the COMBO code
  in corresponding setups, although for the COMBO-code-setups usually
  less objects are rejected. \emph{BPZ} produces $f_{3\sigma}$ values
  which are not too different from the COMBO code. However, one should
  mention at this point that \emph{Hyperz} gives the user a handle to
  get rid of some of these outliers with the drawback of decreased
  completeness. As described in Sect.~\ref{sec:rejection} the outlier
  rate for most \emph{Hyperz} setups decreases monotonically at least
  down to some point when objects with very large photo-$z$ error
  estimates are excluded.}

\begin{figure*}
\centering
\includegraphics[ angle=-90, width=17cm]{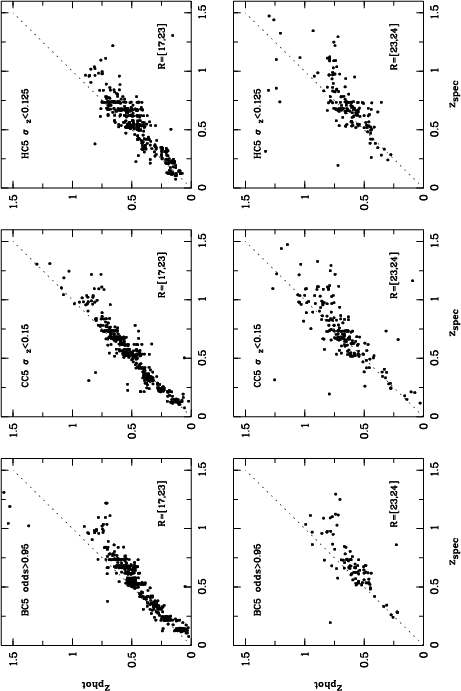}
\caption{\label{fig:zz_BCH_C5}Photo-$z$'s from the CDFS-COMBO $UBVRI$
  imaging data vs. spectroscopic redshifts from the VVDS. The
  \emph{left} column shows results for \emph{BPZ}, the \emph{middle}
  column for the COMBO code, and the \emph{right} column for
  \emph{Hyperz}. Bright objects with $17<R<23$ are shown in the
  \emph{top} panel, faint objects with $23<R<24$ in the \emph{bottom}
  panel.}
\end{figure*}

\begin{figure*}
\resizebox{\hsize}{!}{\includegraphics[angle=-90]{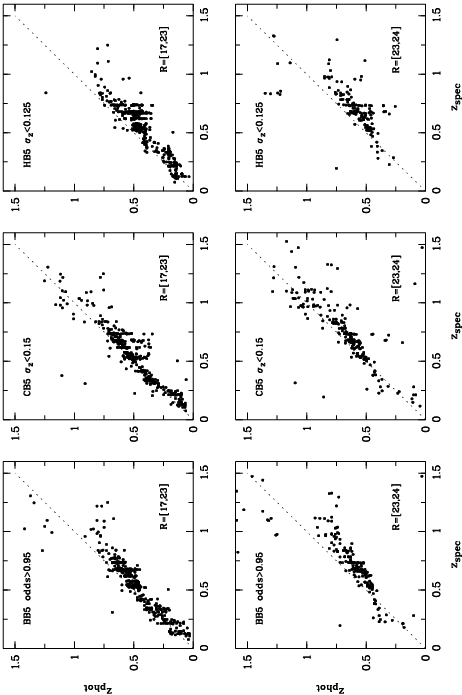}}
\caption{\label{fig:zz_BCH_B5}Same as Fig.~\ref{fig:zz_BCH_C5} but for
  the GaBoDS $UBVRI$ imaging data.}
\end{figure*}

The outlier-excluded scatter values, $\sigma_z$, do not show a clear
trend with every code being the most accurate in at least one setup.
There is clearly some amount of degeneracy between completeness,
$f_{3\sigma}$, $\delta_z$, and $\sigma_z$. The plots in
Fig.~\ref{fig:zz_BCH_C5} and \ref{fig:zz_BCH_B5} provide a more
complete view of the performance.

Remarkably, the negative biases introduced by \emph{BPZ} and
\emph{Hyperz} as reported above are much smaller or negligible for the
COMBO code. { This suggests that the consistent photometric
  calibration of the two surveys (note that the photometry is also
  consistent with the MUSYC survey) is not the source of the
  biases. Rather the combination of these ground-based photometric
  datasets with particular template sets seems to be
  problematic. \emph{BPZ} and \emph{Hyperz} together with the supplied
  template sets} are tested in their release papers
\citep{2000ApJ...536..571B, 2000A&A...363..476B} only against real
data from the Hubble Deep Field, besides simulations. \emph{BPZ} now
incorporates a new template set (see Sect.~\ref{sec:bpz}) that was
specially calibrated for HST photometry. The COMBO code, however, was
originally designed for the ground-based survey CADIS
\citep{2001A&A...365..660W}, where colours were measured bias-free
from seeing adaptive photometry, and included photo-$z$'s for
point-source QSOs.

{ In general, photo-z biases can be removed by a recalibration
  procedure with a spectroscopic training sample. Fixing the redshift
  for the training set objects one can fit for zeropoint offsets in
  the different filters that minimise the magnitude differences
  between the observed object colours and best-fit template
  colours. We developed such recalibration methods for \emph{BPZ} and
  \emph{Hyperz} making use of the spectroscopic redshifts of the
  VVDS. In this way we can decrease or completely remove the biases
  which are still present in the blind setups. A more advanced
  technique incorporating also a recalibration of the template set
  after recalibrating the photometric zeropoints can lead to even more
  accurate results \cite[see
    e.g. ][]{2004ApJS..150....1B,2006A&A...457..841I}. We don't refer
  to the recalibrated photometry in the remainder of this paper and
  instead focus on blind applications.}

One of the biggest differences between the codes is the template set
chosen and one might presume that most of the difference in
performance originates from this point. However, we run \emph{Hyperz}
with the PEGASE templates used by the COMBO code as well as with the
CWW templates plus two Kinney starburst templates originally used by
\emph{BPZ} in \cite{2000ApJ...536..571B}. We switch off the
\emph{Hyperz} internal reddening because it is already included in the
\emph{BPZ} templates and the PEGASE age$\times$extinction grid used by
the COMBO code. The results can neither compete with the best
\emph{Hyperz} setups incorporating the BC templates nor with the COMBO
code plus PEGASE templates. Hence, the implementation of user-defined
templates appears to be not straightforward and results may not be
competitive with the template sets that are shipped with the code and
were tested and optimised by the author.

Another interesting point is the comparison of the CC17 setup with the
CB5 setup. While the total exposure time with WFI is lower for CC17,
the performance of CC17 is better in all statistics described here. It
is clear, that for the particular application of photo-$z$'s
for bright objects, the exposure time was well spent on more filters
(which is an important result for future surveys). However, the GaBoDS
data of the CDFS are completely based on archive data and no specific
observing programme was proposed to create these deep images.
Furthermore, for deeper applications, such as Lyman-break galaxy
studies, where you simply need a very deep colour index between three
bands, the GaBoDS data are certainly highly superior to the COMBO
data.

\begin{table*}
  \caption{Photo-$z$ errors and outlier rates for selected setups on the COMBO-CDFS data (bright sample \emph{left}, faint sample \emph{right}).}
\label{tab:res_VVDS_COMBO}
\begin{tabular}{l r r r r r r r r}
  \hline
  \hline
  Sample & \multicolumn{4}{c}{$R=[17,23]$} & \multicolumn{4}{c}{$R=[23,24]$} \\
  Mean redshift & \multicolumn{4}{c}{0.55} & \multicolumn{4}{c}{0.75} \\
  \hline
  Configuration & compl. [\%] & $f_{0.15}$ [\%] & $f_{3\sigma}$ [\%] & $\left<\delta_z\right>\pm\sigma_z$ & compl. [\%] & $f_{0.15}$ [\%] & $f_{3\sigma}$ [\%] & $\left<\delta_z\right>\pm\sigma_z$\\
  \hline
  \hline
  BC17   &  97.6 &  2.7 & 3.4 & $-0.035\pm0.034$ &  69.7 &  4.6 &  4.6 & $-0.036\pm0.038$ \\
  BC5    &  94.1 &  3.3 & 4.5 & $-0.049\pm0.037$ &  42.2 &  8.7 &  5.4 & $-0.046\pm0.053$ \\
  BC4    &  85.3 &  6.9 & 9.2 & $-0.033\pm0.048$ &  34.4 &  8.0 &  6.7 & $-0.040\pm0.046$ \\
  \hline                                                                    
  CC17   &  99.8 &  1.2 & 4.8 & $-0.011\pm0.018$ & 100.0 &  9.6 & 16.1 & $-0.012\pm0.027$ \\
  CC5    & 100.0 &  4.0 & 5.7 & $-0.017\pm0.035$ &  98.6 &  8.4 &  4.7 & $-0.024\pm0.065$ \\
  CC4    &  97.2 &  7.6 & 5.9 & $-0.023\pm0.056$ &  83.5 & 21.4 & 13.2 & $ 0.001\pm0.084$ \\
  \hline                                                                    
  HC17   &  99.8 &  5.2 & 5.7 & $-0.026\pm0.041$ &  99.5 & 16.6 & 16.1 & $-0.032\pm0.052$ \\
  HC5    &  96.9 &  6.8 & 3.7 & $-0.045\pm0.053$ &  83.5 & 15.9 &  6.6 & $-0.045\pm0.072$ \\
  HC4    &  81.3 & 10.5 & 7.0 & $-0.034\pm0.059$ &  73.4 & 16.9 & 10.0 & $-0.049\pm0.063$ \\
  \hline
\end{tabular}
\end{table*}

\begin{table*}
  \caption{Same as Table~\ref{tab:res_VVDS_COMBO} but for the GaBoDS-CDFS data.}
\label{tab:res_VVDS_GaBoDS}
\begin{tabular}{l r r r r r r r r}
  \hline
  \hline
  Sample & \multicolumn{4}{c}{$R=[17,23]$} & \multicolumn{4}{c}{$R=[23,24]$} \\
  Mean redshift & \multicolumn{4}{c}{0.55} & \multicolumn{4}{c}{0.75} \\
  \hline
  Configuration & compl. [\%] & $f_{0.15}$ [\%] & $f_{3\sigma}$ [\%] & $\left<\delta_z\right>\pm\sigma_z$ & compl. [\%] & $f_{0.15}$ [\%] & $f_{3\sigma}$ [\%] & $\left<\delta_z\right>\pm\sigma_z$\\
  \hline
  \hline
  BB5    &  97.4 &  2.7 &  3.4 & $-0.037\pm0.040$ &  87.7 &  6.7 & 12.4 & $-0.062\pm0.041$ \\
  BB4    &  88.5 &  3.1 & 10.4 & $-0.010\pm0.045$ &  85.0 &  5.3 &  9.6 & $-0.060\pm0.043$ \\
  \hline                                                   
  CB5    &  99.5 &  3.6 &  4.3 & $-0.024\pm0.041$ &  98.6 &  8.8 &  8.8 & $-0.028\pm0.044$ \\
  CB4    &  99.8 &  4.8 &  3.8 & $-0.019\pm0.049$ &  96.8 & 11.4 & 12.3 & $-0.026\pm0.042$ \\
  \hline                                                   
  HB5    &  95.7 &  9.0 &  3.8 & $-0.065\pm0.048$ &  79.1 & 18.4 & 19.0 & $-0.060\pm0.039$ \\
  HB4    &  74.2 &  4.8 &  5.5 & $-0.034\pm0.040$ &  75.5 & 13.3 & 13.9 & $-0.052\pm0.035$ \\
  \hline
\end{tabular}
\end{table*}

\subsection{FDF results}
\begin{figure*}
\centering
\includegraphics[ angle=-90, width=17cm]{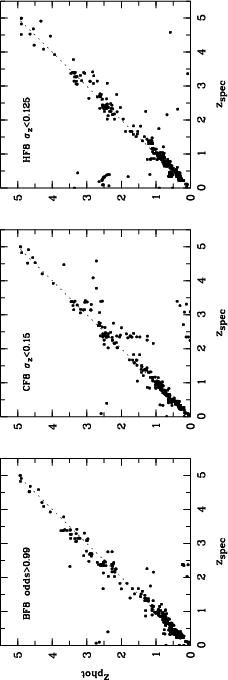}
\caption{\label{fig:FDF}Photometric vs. spectroscopic redshifts for
  the FDF full 8-filter set imaging data. The \emph{left} diagram
  shows results for \emph{BPZ}, the \emph{middle} diagram for the
  COMBO code, and the \emph{right} diagram for \emph{Hyperz}.}
\end{figure*}

Table~\ref{tab:res_FDF} summarises the results on the FDF and
Fig~\ref{fig:FDF} shows photometric vs. spectroscopic redshift for
selected setups. In the lower redshift bin again the COMBO code
combined with imaging data in 8 filters delivers the smallest outlier
rate, bias, and scatter when compared to \emph{BPZ} and \emph{Hyperz}
in 8 filters. At least in this redshift interval the results are
nearly as good as the results produced by \cite{2004A&A...421...41G}
with a template set specifically calibrated for the FDF.

In the high redshift domain, however, the COMBO code does not perform
well with an outlier rate and scatter twice as large as the ones
produced by \emph{Hyperz} and with a considerable bias. \emph{BPZ}
performs not too different from \emph{Hyperz}. Apparently, the COMBO
code in combination with the PEGASE templates has problems when the
Lyman break enters the filter set: many objects appear at too low
redshifts, hence the large negative bias (see also Fig~\ref{fig:FDF}).
The inferior performance in the high redshift domain can then be
attributed to colour-redshift-degeneracies described in detail in
\cite{2000ApJ...536..571B}. Basically, a larger number of templates
can lead to better low-$z$ performance with the tradeoff of poorer
high-$z$ performance due to increasing degeneracies.  Designed for
medium-deep surveys the COMBO code was naturally not optimised to work
at high redshifts in contrast to \emph{BPZ} and \emph{Hyperz}. There,
the application of a Bayesian prior on the apparent magnitude combined
with a sparse template set (\emph{BPZ}) or a top-hat prior on the
absolute magnitude (\emph{Hyperz}) delivers significantly better
results.

The dependence of photo-$z$ performance on the filter set
is also shown in Table~\ref{tab:res_FDF}. In the lower redshift
interval the outlier rate nearly doubles as soon as the NIR filters $J$
and $K_s$ are dropped. The scatter, however, remains nearly constant.
Without near-infrared data a larger negative bias is introduced which
was already present in all VVDS-\emph{Hyperz} setups (see
Table~\ref{tab:res_VVDS_COMBO}~and~\ref{tab:res_VVDS_GaBoDS}). The
exclusion of the peculiar $U$-band reduces this bias again with the
drawback of increased scatter. In the higher redshift domain results
get much worse when near-infrared data are dropped.

\begin{table*}
  \caption{Same as Tables~\ref{tab:res_VVDS_COMBO} and \ref{tab:res_VVDS_GaBoDS} but for the FDF data (low-$z$ sample \emph{left}, high-$z$ sample \emph{right}).}
\label{tab:res_FDF}
\begin{tabular}{l r r r r r r r r}
\hline
\hline
Sample & \multicolumn{4}{c}{$z=[0,2]$} & \multicolumn{4}{c}{$z=[2,5]$} \\
Mean redshift & \multicolumn{4}{c}{0.65} & \multicolumn{4}{c}{2.94} \\
\hline
  Configuration & compl. [\%] & $f_{0.15}$ [\%] & $f_{3\sigma}$ [\%] & $\left<\delta_z\right>\pm\sigma_z$ & compl. [\%] & $f_{0.15}$ [\%] & $f_{3\sigma}$ [\%] & $\left<\delta_z\right>\pm\sigma_z$\\
\hline
BF\_UBgRIZJKs &  98.1 &  5.4 &  5.4 & $ 0.005\pm0.053$ &  93.3 & 12.0 & 13.3 & $ 0.026\pm0.046$ \\
CF\_UBgRIZJKs &  99.6 &  3.4 &  4.9 & $ 0.001\pm0.034$ & 100.0 & 29.2 & 13.5 & $-0.046\pm0.093$ \\
HF\_UBgRIZJKs &  99.6 & 10.2 &  9.8 & $-0.019\pm0.051$ & 100.0 &  8.0 &  5.7 & $-0.004\pm0.056$ \\
\hline                                                           
BF\_UBGRIJKs  &  98.5 &  8.4 &  9.2 & $ 0.011\pm0.058$ &  89.9 & 18.8 & 18.8 & $ 0.024\pm0.050$ \\
BF\_UBGRIZ    &  97.7 &  7.7 &  9.6 & $-0.010\pm0.041$ &  91.0 & 17.3 & 19.8 & $ 0.032\pm0.047$ \\
BF\_UBGRI     &  98.1 &  8.4 &  9.2 & $ 0.000\pm0.042$ &  74.2 & 27.3 & 27.3 & $ 0.025\pm0.057$ \\
BF\_BGRI      &  90.2 & 12.5 &  8.3 & $ 0.020\pm0.060$ &  53.9 & 27.1 & 27.1 & $ 0.017\pm0.051$ \\
\hline
CF\_UBGRIJKs  &  99.6 &  4.5 &  4.5 & $ 0.006\pm0.044$ &  95.5 & 30.6 & 14.1 & $-0.058\pm0.102$ \\
CF\_UBGRIZ    &  99.6 &  6.0 &  7.9 & $-0.006\pm0.039$ &  98.9 & 54.5 & 47.7 & $-0.061\pm0.077$ \\
CF\_UBGRI     &  96.6 &  5.8 &  7.0 & $-0.001\pm0.043$ &  94.4 & 57.1 & 38.1 & $-0.100\pm0.082$ \\
CF\_BGRI      &  92.5 & 16.3 & 13.0 & $ 0.006\pm0.066$ &  93.3 & 62.7 & 50.6 & $-0.095\pm0.099$ \\
\hline
HF\_UBGRIJKs  &  99.6 & 10.2 & 10.6 & $-0.022\pm0.049$ & 100.0 &  9.1 &  8.0 & $-0.012\pm0.053$ \\
HF\_UBGRIZ    & 100.0 & 12.8 & 11.7 & $-0.020\pm0.048$ & 100.0 & 12.5 & 12.5 & $ 0.009\pm0.052$ \\
HF\_UBGRI     &  99.2 & 14.4 & 17.8 & $-0.028\pm0.040$ &  97.7 & 20.9 & 14.0 & $-0.012\pm0.072$ \\
HF\_BGRI      &  99.6 & 19.6 & 22.6 & $-0.021\pm0.038$ &  96.6 & 24.7 & 16.5 & $-0.010\pm0.086$ \\
\hline
\end{tabular}
\end{table*}

{ 
\section{Photo-$z$ vs. photo-$z$ comparisons}
\label{sec:phz_vs_phz}
With photo-$z$ estimates for the complete imaging catalogues at hand
we cannot only compare photo-$z$'s to spectroscopic redshifts but we
can also compare the different photo-$z$'s to each other. In this way
we are able to detect possible selection effects that might still be
present in the secure spectroscopic subsamples used in the preceding
sections.

We define similar quantities as in Sect.~\ref{sec:strategy} but now
with the spectroscopic redshift replaced by another photo-$z$. Since
none of the two photo-$z$'s is superior to the other in general, the
interpretation of the statistics changes then. For example, a
catastrophic disagreement between two photo-$z$ estimates just means
that at least one of the two is wrong, but it can also be true that
both are wrong.

Due to these complications we can learn most from comparing the
photo-$z$ vs. photo-$z$ benchmarks for different subsamples. In the
following, we will look at the complete CC5 catalogue and compare
these redshift estimates with CB5. Moreover we compare CB5 to
HC5. Thus, we study how the performance is either affected by
additional depth or by using a different code. Two samples are
considered, the whole catalogue with $17<R<23$ and the subsample with
secure spectroscopic redshifts used before. Any significant
statistical difference in these photo-$z$ vs. photo-$z$ comparisons
can be interpreted then to be due to selection effects in the
spectroscopic subsample.

Figures~\ref{fig:phz_vs_phz_depth}~\&~\ref{fig:phz_vs_phz_codes} show
the results for a comparison of photo-$z$'s from data of different
depths and from different codes, respectively. The statistics are
summarised in Table~\ref{tab:phz_vs_phz}. We require an object to meet
the criteria defined in Sect.~\ref{sec:rejection} for both photo-$z$
setups entering the comparison.

There are some distinctive features visible in
Fig.~\ref{fig:phz_vs_phz_depth}. The ones labelled ``A'' and ``B'' can
be found in in both panels and just the overall number density in the
left panel is larger by a factor of 14. Two other features, one clump
at $z_{\mathrm phot,CC5}=0.0-0.3$ and $z_{\mathrm phot,CB5}\sim0.4$
and a couple of outlier objects at $z_{\mathrm phot,CC5}=0.5-1.5$ and
$z_{\mathrm phot,CB5}=0.0-0.1$, are however, only found in the
photometric sample and not in the spectroscopic one. 

Even more striking is the difference in the overall distributions of
objects when the two codes are compared in
Fig.~\ref{fig:phz_vs_phz_codes}.

This is also reflected in the numbers. The outlier rates, $f_{0.15}$,
for the full sample are larger by a factor of $\sim2$ when comparing
data depth (CC5 vs. CB5) and $\sim4$ when comparing codes on identical
data (CB5 vs. HB5). Completeness and scatter are essentially the same
for both subsamples while the bias slightly increases from ``CB5
vs. HB5, all'' to ``CB5 vs. HB5, spectro''.

This means that the properties of the core of the $\Delta z$
distribution are quite similar for both samples but that the wings are
more pronounced when all objects are considered. Apparently, the
secure spectroscopic subsample represents an intrinsically different
galaxy population than the full, purely magnitude-limited sample. The
rejection of objects with bad spectroscopic flags introduces a bias in
the spectroscopic sample so that it is no longer purely
magnitude-limited and artificially reduces the apparent outlier rates.

\begin{figure*}
%\sidecaption
\includegraphics[width=12cm]{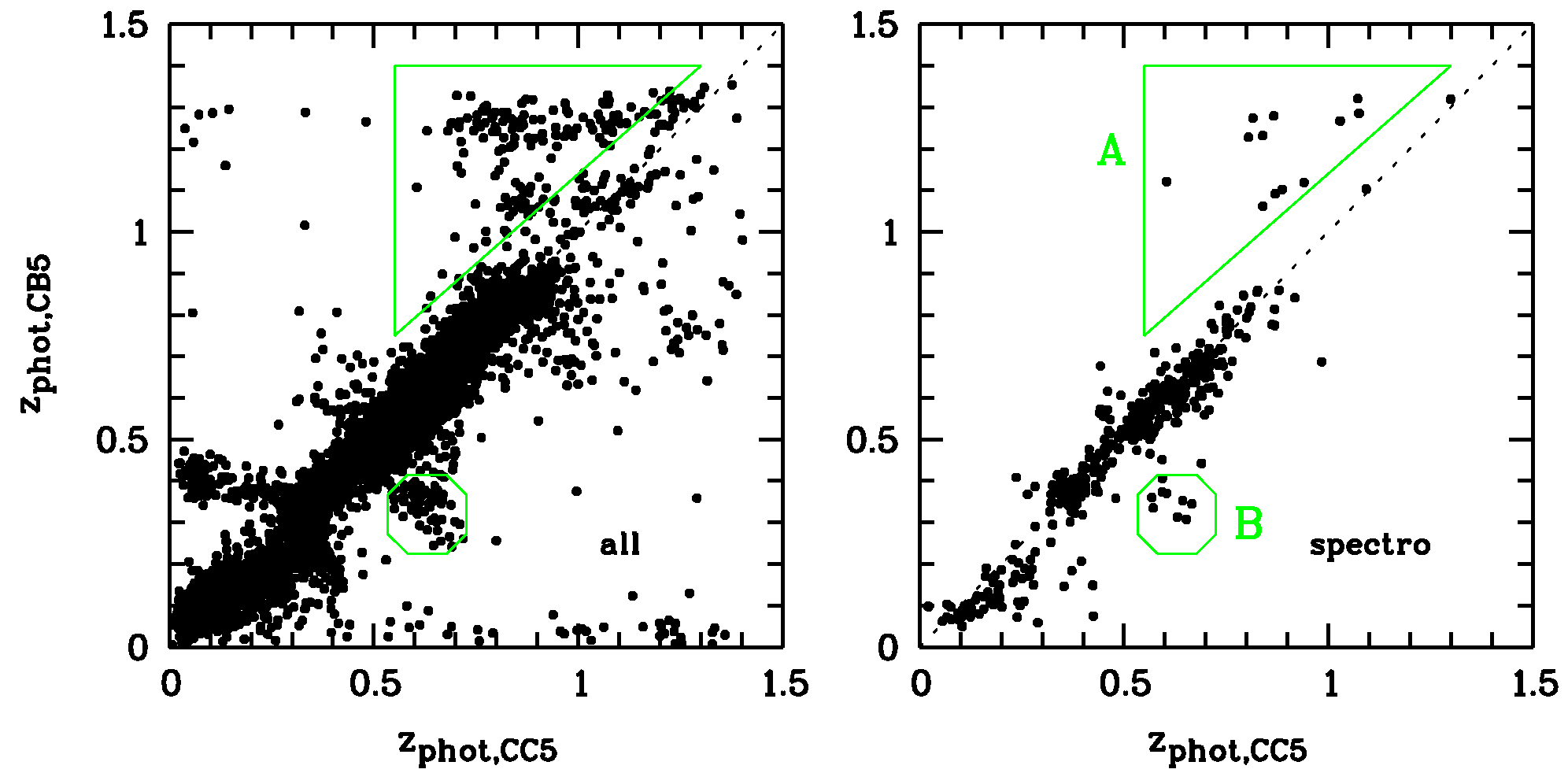}
\caption{\label{fig:phz_vs_phz_depth}Photo-$z$ vs. photo-$z$ for the
  COMBO code run on the $UBVRI$ imaging data from COMBO (CC5) and from
  GaBoDS (CB5). The \emph{left} diagram shows results for the full
  sample and the \emph{right} diagram for the secure spectroscopic
  subsample. In this plot the objects rejected by the criteria from
  Sect.~\ref{sec:rejection} are not shown. The object density in the
  \emph{left} plot is 14 times higher than in the \emph{right} one.}
\end{figure*}

\begin{figure*}
%\sidecaption
\includegraphics[width=12cm]{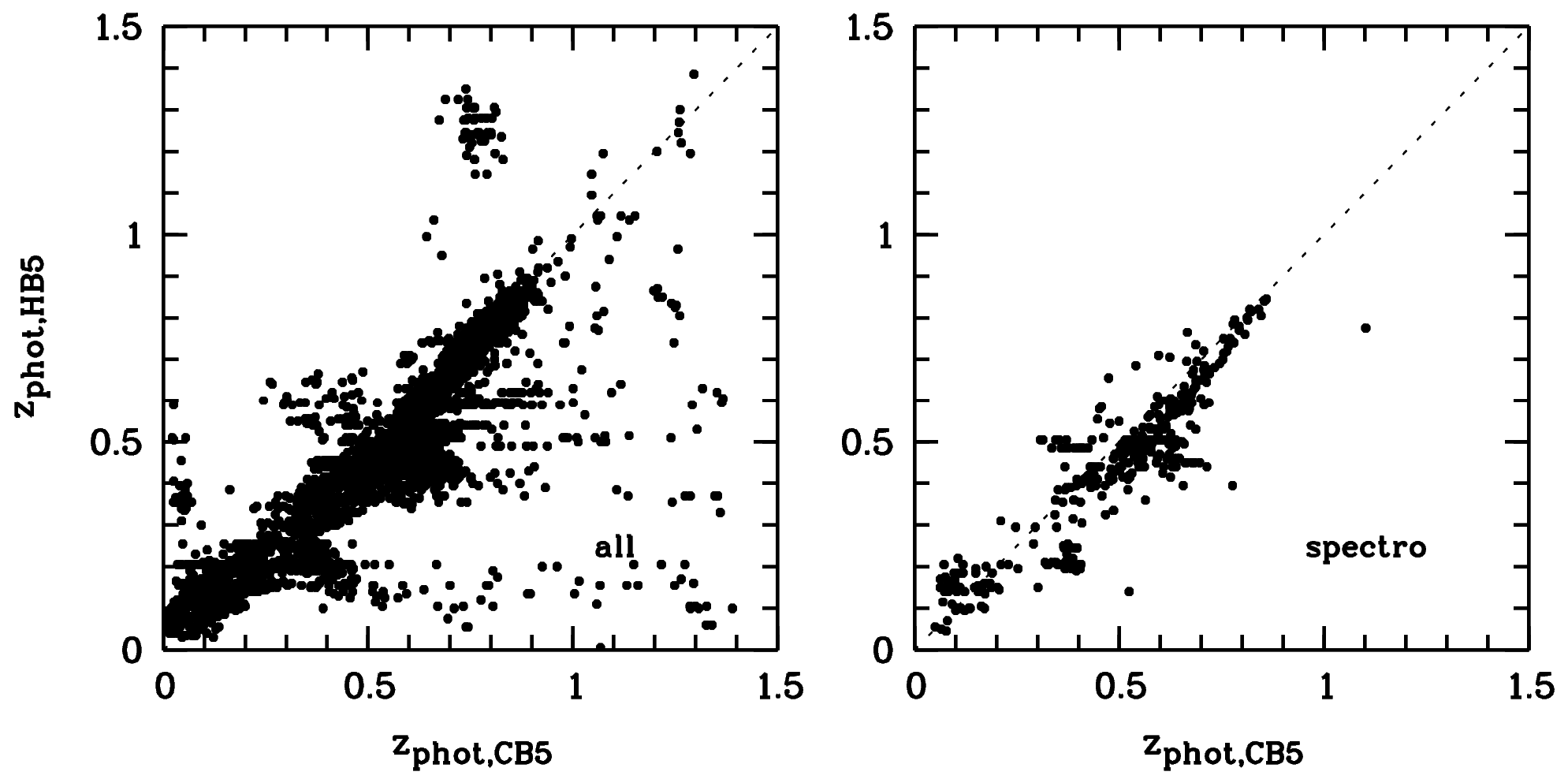}
\caption{\label{fig:phz_vs_phz_codes}Same as
  Fig.~\ref{fig:phz_vs_phz_depth} but with photo-$z$ vs. photo-$z$ for
  the COMBO code (CB5) and \emph{Hyperz} (HB5) run on the GaBoDS
  $UBVRI$ imaging data.}
\end{figure*}

\begin{table*}
  \caption{Statistics for the comparison between the different photo-$z$'s.}
\label{tab:phz_vs_phz}
\begin{tabular}{l r r r r}
  \hline
  \hline
  Sample & \multicolumn{4}{c}{$R=[17,23]$}\\
  \hline
  Configuration & compl. [\%] & $f_{0.15}$ [\%] & $f_{3\sigma}$ [\%] & $\left<\delta_z\right>\pm\sigma_z$\\
  \hline
  \hline
  CC5 vs. CB5, all     & 99.4 & 6.3 & 9.7 & $-0.007\pm0.035$\\
  CC5 vs. CB5, spectro &100.0 & 3.2 & 9.8 & $-0.006\pm0.030$\\
  \hline
  CB5 vs. HB5, all     & 92.6 & 9.0 & 8.7 & $-0.033\pm0.053$\\
  CB5 vs. HB5, spectro & 95.7 & 2.5 & 2.8 & $-0.057\pm0.047$\\
  \hline
\end{tabular}
\end{table*}
}

\section{Summary and Conclusions}
\label{sec:conclusions}
We have shown that photo-$z$'s estimated with today's tools can
produce a reasonable accuracy. The performance of a particular
photo-$z$ code, however, cannot easily be characterised by a mere two
numbers such as scatter and global outlier rate. The benchmarks are
rather sensitive functions of filter set, depth, redshift range and
code settings.  Moreover, there is at least a factor of two possible
difference in performance between different codes which is again not
stable for all setups but can vary considerably from one setup to
another. There are, for example, redshift ranges where one code
clearly beats another one in terms of accuracy only to loose at other
redshifts. We give estimates of the performance for a number of codes
in some practically relevant cases.

The estimation of photo-$z$'s from different ground-based datasets is
not straightforward and results should not be expected to be identical
to simulated photo-$z$ estimates. Rather, photo-$z$ simulations often
seem to circumvent critical steps in ground-based photo-$z$
estimation. { Most importantly}, the match between observed colours
and some template sets commonly used may be suboptimal.

In the preceding sections we have identified several aspects which are
relevant to future optimisations of photo-$z$ codes. The photo-$z$
error estimation is one of the most unsatisfying aspects to date with
error values often only very weakly correlated with real
uncertainties. This is likely due to the insufficient inclusion of
systematics since very low S/N objects, for which the errors should be
dominated by photon shot-noise, show a tighter correlation.
Chip-to-chip sensitivity variations, especially in the UV, could
either be taken into account more accurately within the photo-$z$
codes or could be tackled by improved instrument design, survey
strategy, and data reduction. The optimisation of template sets can be
expected to be successfully done with ever larger spectroscopic
catalogues becoming available.

{ In general, biases can be removed by a recalibration which
  requires an extensive spectroscopic training set. Another proven
  successful route to better photo-$z$'s is improving the spectral
  resolution of the data, instead of their depth, as demonstrated by
  the COMBO-17 survey. This approach is also taken by the new ALHAMBRA
  survey \citep[][ Ben\'itez et al., 2007, A\&A,
    submitted]{2005astro.ph..4545M} and COSMOS-21.

A general problem for all studies comparing photo-$z$'s to
spectroscopic redshifts is our finding that secure spectroscopic
samples can be biased. While surveys like VVDS are $>90\%$ complete in
obtaining spectra for galaxy samples the redshifts that are claimed to
be $>90\%$ secure only amount to $\sim50\%$. This subsample obviously
consists of galaxies for which the photo-$z$ estimation works better
than for the whole sample. In the future, it is desirable to put
effort into spectroscopic surveys with secure redshift measurements
for virtually every galaxy down to the same flux limit that is used
for the analysis of photo-$z$ samples.

Several questions that are raised in this work will be tackled by the
PHAT initiative mentioned above. PHAT aims to understand the issues
presented here in a systematical and quantitative way in order to give
guidance for better photo-$z$'s in the future.}

\begin{acknowledgements}
This work was supported by the German Ministry for Education and
Science (BMBF) through the DLR under the project 50 OR 0106, by the
BMBF through DESY under the project 05 AV5PDA/3, and by the Deutsche
Forschungsgemeinschaft (DFG) under the projects SCHN342/3-1 and
ER327/2-1. CW was supported by a PPARC Advanced Fellowship.
\end{acknowledgements}

\bibliographystyle{aa}

\bibliography{photo_z_2007}

\end{document}